\documentclass[twoside,draft,reqno]{birkart}
\usepackage{amssymb}
\input epsf

\renewcommand{\theequation}{\arabic{section}.\arabic{equation}}
\newcommand{\news}{\setcounter{equation}{0}}
\newcommand{\be}{\begin{equation}}
\newcommand{\ee}{\end{equation}}
\newcommand{\bea}{\begin{eqnarray}}
\newcommand{\eea}{\end{eqnarray}}
\newcommand{\bean}{\begin{eqnarray*}}
\newcommand{\eean}{\end{eqnarray*}}
\font\upright=cmu10 scaled\magstep1
\font\sans=cmss12
\newcommand{\ssf}{\sans}
\newcommand{\stroke}{\vrule height8pt width0.4pt depth-0.1pt}
\newcommand{\Z}{\hbox{\upright\rlap{\ssf Z}\kern 2.7pt {\ssf Z}}}

\newcommand{\C}{{\rlap{\rlap{C}\kern 3.8pt\stroke}\phantom{C}}}
\newcommand{\R}{\hbox{\upright\rlap{I}\kern 1.7pt R}}
\newcommand{\CP}{\C{\upright\rlap{I}\kern 1.5pt P}}

\newcommand{\identity}{{\upright\rlap{1}\kern 2.0pt 1}}
\newcommand{\bm}{\boldmath}
\newcommand{\sk}{Skyrmion}
\newcommand{\Sk}{Skyrme}
\newcommand{\I}{{\calI}} 
\newcommand{\calO}{{\mathcal O}} 
\newcommand{\calI}{{\mathcal I}} 

\begin{document}
\setcounter{page}{1}

\pagestyle{plain}

\title{\vskip -70pt
\begin{flushright}{\normalsize DAMTP 00-???} \\
\end{flushright}\vskip 20pt
{\bf \Large \bf Understanding Skyrmions using Rational Maps}\vskip 10pt}
\title[Understanding Skyrmions using Rational Maps]
{ Understanding Skyrmions using Rational Maps
}
\author[N.~Manton and B.M.A.G.~ Piette]{Nicholas S. Manton$^{\dagger}$ 
and Bernard M.A.G. Piette$^{\ddagger}$}
\address{
$\dagger$ Department of Applied Mathematics and Theoretical Physics,\br
University of Cambridge,\br 
Wilberforce Road, Cambridge CB3 0WA, England.}
\email{N.S.Manton@damtp.cam.ac.uk}
\address{
$\ddagger$ Department of Mathematical Sciences,\br
University of Durham,\br
South Road, Durham DH1 3LE, England}
\email{B.M.A.G.Piette@durham.ac.uk}

\begin{abstract}
We discuss an ansatz for Skyrme fields in three dimensions which
uses rational maps between Riemann spheres, and produces shell-like
structures of polyhedral form. Houghton, Manton and Sutcliffe showed that
a single rational map gives good approximations to the minimal energy 
Skyrmions up to baryon number of order 10. We show how the method can be
generalized by using two or more rational maps to give a double-shell
or multi-shell structure. Particularly interesting examples occur at
baryon numbers 12 and 14.
\end{abstract}
\maketitle

\section{Introduction}

The Skyrme model is a nonlinear theory of pions in $\R^3$, 
with an $SU(2)$ valued scalar field $U({\bf x},t)$,
the Skyrme field, satisfying the boundary condition
$U\rightarrow 1$ as ${ \vert {\bf x}\vert \rightarrow\infty}.$ Static
fields obey the equation
\begin{equation}
\partial_i(R_i-\frac{1}{4}[R_j,[R_j,R_i]])=0
\label{sky}
\end{equation}
where $R_i$ is the $su(2)$ valued current $R_i=(\partial_iU)U^{-1}.$
Such fields are stationary points (either minima or saddle points) of the 
energy function
\begin{equation}
E=\int \{-\frac{1}{2}\mbox{Tr}(R_iR_i)-\frac{1}{16}
\mbox{Tr}([R_i,R_j][R_i,R_j])\} \ d^3{\bf x}.
\label{energy}
\end{equation}

Associated with a Skyrme field is a topological integer, the baryon
number $B$, defined as the degree of the map $U: \R^3\mapsto SU(2)$.
It is
well defined because of the boundary condition at infinity.

Solutions of eq. (\ref{sky}) are known for several values of $B$,
but they can only be obtained numerically. Many of these solutions are
stable, and probably represent the global minimum of the energy for
given $B$. We shall refer to the solutions believed to be of lowest
energy for each $B$ as \sk s.

There is a nine-dimensional symmetry group of the
equation and boundary condition. It consists of translations 
and rotations in $\R^3$
and the $SO(3)$ isospin transformations $U\mapsto {\calO}
U{\calO}^{-1}$ where ${\calO}$ is a constant element of $SU(2)$. 

Skyrme found the spherically symmetric $B=1$ \sk. The $B=2$ \sk\ is toroidal. 
A substantial numerical search for \sk\ solutions was undertaken
by Braaten, Townsend and Carson \cite{BTC}, and minimal energy 
solutions up to $B=5$
were found.  (Their solution for $B=6$ was rather inaccurate.) 
Surprisingly, the $B=3$
solution has tetrahedral symmetry $T_d$, and the $B=4$ solution has
cubic symmetry $O_h.$  Battye and Sutcliffe \cite{BS2} subsequently
found all \sk s up to $B=8$. (Their solution for $B=9$ is probably a
saddle point.) The $B=7$ \sk\ has icosahedral symmetry $Y_h$.  Recently,
with new methods, they have found candidate solutions up to $B=22$ 
\cite{BS4}.

\section{Skyrme Fields from Rational Maps}
\news

Let us denote a point in $\R^3$ by its coordinates $(r,z)$ where $r$
is the radial distance from the origin and 
$z = \tan(\theta/2)\exp i\varphi$ specifies the direction
from the origin. Houghton, Manton and Sutcliffe \cite{HMS} showed that 
one can understand the structure of the known Skyrmions 
in terms of an ansatz using rational maps 
\be 
R(z)=\frac{p(z)}{q(z)} \label{rat}
\end{equation} 
where $p$ and $q$ are polynomials, and a radial profile function
$f(r)$. One identifies 
the target $S^2$ of the rational maps with spheres of latitude on
$SU(2)$, that is, spheres at a fixed distance from the identity element. 

Recall that the direction $z$ corresponds to the Cartesian unit vector 
\begin{equation}
\widehat{\bf n}_z=\frac{1}{1+\vert
  z\vert^2}(2{\rm Re}(z), 2{\rm Im}(z),1-\vert z\vert^2).
\label{unit1}
\end{equation}
Similarly the value of the rational map $R$ is associated with the unit vector

\begin{equation}
\widehat{\bf n}_R=\frac{1}{1+\vert R\vert^2}
(2{\rm Re}(R), 2{\rm Im}(R),1-\vert R\vert^2).
\label{unit2}
\end{equation}
The ansatz is
\begin{equation}
U(r,z)=\exp(if(r) \ \widehat{\bf n}_{R(z)}\cdot\mbox{\bm $\sigma$})
\label{ansatz}
\end{equation}
where $\mbox{\bm $\sigma$}=(\sigma_1,\sigma_2,\sigma_3)$ are Pauli matrices.
For this to be well defined at the origin, $f(0)=k\pi$, for some integer $k.$
The boundary value $U=1$ at $r=\infty$ requires that $f(\infty)=0.$ 
The baryon number of this field is $B=Nk$, where 
$N={\rm max}({\rm deg}\ p, {\rm deg}\ q)$ is the degree of $R.$
We consider only the case $k=1$ here,
so $B=N.$ Note that an $SU(2)$  M\"obius
transformation of the rational map 
\begin{equation}
  R(z)\mapsto \frac{\alpha
  R(z)+\beta}{-\bar\beta R(z)+\bar\alpha} 
\label{mobius}
\end{equation}
with $\vert\alpha\vert^2+\vert\beta\vert^2=1$ acts as an isospin 
transformation.

An attractive feature of the ansatz (\ref{ansatz}) is that it leads to
a simple energy expression which can be minimized with respect
to the rational map $R$ and the profile function $f$ to obtain close
approximations to the true \sk s. A generalized rational map ansatz
has also proved useful in the construction of solutions to Skyrme
models with fields having values in $SU(n)$ \cite{IPZ}.

The energy for a Skyrme field of the form (\ref{ansatz}) is
\begin{eqnarray}
E=\int\bigg[
{f'}^2 +2({f'}^2 + 1)\frac{\sin^2f}{r^2}
 \bigg(\frac{1+\vert z\vert^2}{1+\vert R\vert^2}
       \bigg\vert\frac{dR}{dz}\bigg\vert\bigg)^2 \nonumber \\
   +\frac{\sin^4f}{r^4}
     \bigg(
       \frac{1+\vert z\vert^2}{1+\vert R\vert^2}
           \bigg\vert\frac{dR}{dz}\bigg\vert\bigg)^4\bigg]
   \frac{2i \  dzd\bar z r^2  dr}{(1+\vert z\vert^2)^2}. 
\label{energy3}
\end{eqnarray}
Now
\begin{equation}
\bigg(
\frac{1+\vert z\vert^2}{1+\vert R\vert^2}
\bigg\vert\frac{dR}{dz}\bigg\vert\bigg)^2
 \frac{2i \   dz  d\bar z }{(1+\vert z\vert^2)^2}
\label{winding}
\end{equation}
is the pull-back of the area form $2i\ dRd\bar R/(1+\vert R\vert^2)^2$
on the target sphere of the rational map; therefore its integral 
is $4\pi$ times
the degree $N$. So the energy simplifies to
\begin{equation}
E=4\pi\int \bigg(
r^2 {f'}^2+2N({f'}^2+1)\sin^2 f+\I\frac{\sin^4 f}{r^2}\bigg) \ dr
\label{energy4}
\end{equation}
where $\I$ denotes the integral
\begin{equation}
\I=\frac{1}{4\pi}\int
 \bigg(
\frac{1+\vert z\vert^2}{1+\vert R\vert^2}
\bigg\vert\frac{dR}{dz}\bigg\vert\bigg)^4 \frac{2i \  dz  d\bar z 
}{(1+\vert z\vert^2)^2}.
\label{i}
\end{equation}

Applying a Bogomolny-type argument to the expression
(\ref{energy4}), one can show that 
\begin{equation}
E\ge 4\pi^2(2N+\sqrt{\I}).
\label{bound}
\end{equation} 
This lower bound on the energy, which applies to the rational map
ansatz, is higher than the Fadeev-Bogomolny bound satisfied by any
Skyrme field, $E \ge 12\pi^2 B$. This is because the Schwarz inequality
implies $\I \ge N^2$.

To minimize the energy $E$, one should first
minimize $\I$ over all maps of degree $N$. Then
the profile function $f$ 
minimizing (\ref{energy4}) is found by solving a second
order differential equation with $N$ and $\I$ as parameters. 
In \cite{HMS} only rational maps of 
a given symmetric
form were considered, with symmetries corresponding to a known Skyrmion.
If these symmetric maps still contained a few free parameters, $\I$ was
minimized with respect to these, and then $f$ was calculated. A rational map, 
$R:S^2\mapsto S^2$, is symmetric under a
subgroup $G\subset SO(3)$ if there is a set of M\"obius transformation
pairs $\{g,D_g\}$ with $g\in G$ acting on the domain $S^2$ and $D_g$
acting on the target $S^2$, such that 
\begin{equation} R(g(z))=D_gR(z).
\label{inv}
\end{equation}
Some rational maps also possess additional reflection or
inversion symmetry. 
In their recent work, Battye and
Sutcliffe have systematically sought the rational map that minimizes
$\I$, up to $N=22$ \cite{BS4}. This work has confirmed that the 
choice of maps in \cite{HMS}, up to $N=8$, was optimal.

The zeros of the Wronskian
\begin{equation}
W(z)=p'(z)q(z)-q'(z)p(z)
\label{wron}
\end{equation}
of a rational map $R(z)$ give interesting information about the
shape of the corresponding Skyrme field. 
Where $W$ is zero, the derivative $dR/dz$ is
zero, so the baryon density vanishes. The energy density
is also low. The Skyrme
field baryon density contours therefore look like a polyhedron with holes
in the directions given by the $2N-2$ zeros of $W$.

\section{Symmetric Rational Maps}
\news

In this Section, we present the symmetric
rational maps of degrees 1 to 8, determined in Ref. \cite{HMS}.
Table 1 gives the energy of the
resulting approximate Skyrmions,
and also the energy of the true Skyrmions.  All
numerical values for the energies are the real
energies divided by $12\pi^2B$, and hence close to unity. Fig.  1
shows a surface of constant baryon density for most of the approximate
Skyrmions. The true solutions
have very similar shapes \cite{BS2}.

For $B=1$ the basic map is $R(z)=z$, for which the integral $\I =1$, 
and (\ref{ansatz}) reduces to Skyrme's hedgehog field
\begin{equation}
U(r,\theta,\varphi)=\cos f+i\sin f(\sin\theta\cos\varphi\ \sigma_1
+\sin\theta\sin\varphi\ \sigma_2+\cos\theta\ \sigma_3).
\label{hh}
\end{equation}
This is $SO(3)$ invariant, since
$R(g(z))=g(z)$ for any $g \in SU(2)$.
It gives the standard exact spherically symmetric Skyrmion with its usual
profile $f(r)$, and with energy $E=1.232$.

The rational map which gives the toroidal $B=2$
\sk\ is 
\begin{equation}
R(z)=z^2.
\label{g2}
\end{equation}
Using this, one
finds $\I=\pi+8/3$ and after determining the profile $f(r)$ one
obtains $E=1.208$, an energy 3\% higher than that of the true solution.

The $B=3$ \sk\ has tetrahedral symmetry. A rational map
with this symmetry is obtained by imposing
\begin{equation}
R(-z)=-R(z) \hskip 1cm \mbox{,} \hskip 1cm R(1/z)=1/R(z) \,
\label{z2b}
\end{equation} 
\begin{equation}
R\bigg(\frac{iz+1}{-iz+1}\bigg)=\frac{iR(z)+1}{-iR(z)+1}.
\label{s3}
\end{equation} 
This gives the degree 3 maps \be
R(z)=\frac{\sqrt{3}az^2-1}{z(z^2-\sqrt{3}a)}
\label{g3}
\end{equation}
with $a=\pm i.$ Note that
\hbox{$z\mapsto (iz+1)/(-iz+1)$} sends 
\hbox{$0\mapsto 1\mapsto i\mapsto 0$} 
and hence generates the $120^\circ$ rotation cyclically
permuting the Cartesian axes. The sign of $a$ can be changed by the 
$90^\circ$ rotation $z\mapsto iz$. For these maps $\I=13.58$. 
Solving for the profile
$f(r)$, one finds an energy $E=1.184$. 
The Wronskian of maps of the form
(\ref{g3}) is proportional to $z^4 \pm 2\sqrt{3}iz^2+1$, 
a tetrahedral Klein polynomial \cite{Kl}.

The $B=4$ \sk\ has cubic symmetry. The cubically symmetric
rational map of degree 4 is the ratio of tetrahedral
Klein polynomials
\begin{equation}
R(z)=\frac{z^4+2\sqrt{3}iz^2+1}{z^4-2\sqrt{3}iz^2+1}.
\label{g4}
\end{equation}
The $90^\circ$ rotation is a symmetry, because
$R(iz)=1/R(z)$.
Using (\ref{g4}) in the ansatz (\ref{ansatz}) gives an energy $E=1.137.$ 

The $B=5$ \sk\ of minimal energy has symmetry $D_{2d},$ which is
somewhat surprising. A nearby cubically symmetric solution exists but is
a saddle point.
The $D_{2d}$-symmetric family of rational maps is
\begin{equation}
R(z)=\frac{z(z^4+bz^2+a)}{az^4-bz^2+1}
\label{g5}
\end{equation}
with $a$ and $b$ real. 
If $b=0$ then $R(z)$ has $D_4$ symmetry,
the symmetry of a square. There is cubic symmetry if, in addition, 
$a=-5$. This value ensures the $120^\circ$ rotational symmetry (\ref{s3})
and the Wronskian is then proportional to $z^8+14z^4+1$,
the face polynomial of an octahedron.
When $b=0,a=-5$, the integral
$\I=52.05$. However, $\I$ is minimized when $a=3.07$, $b=3.94$,
taking the value $\I=35.75.$ This is consistent with the
structure of the $B=5$ \sk,
a polyhedron made from four pentagons and four quadrilaterals.  
With the optimal profile function $f(r)$, the energy 
is $E=1.147$. 
The octahedral saddle point
has $E=1.232$. There is a further, 
much higher saddle point at
$a=b=0$, where the map (\ref{g5}) simplifies to $R(z) = z^5$,
and gives a toroidal field.

The $B=6$ \sk\ has symmetry $D_{4d}$. The rational maps
\begin{equation}
R(z)=\frac{z^4-a}{z^2(az^4+1)}
\label{g6}
\end{equation}
have this symmetry, and the minimal energy occurs at $a=0.16$, 
giving $E=1.137.$
The Skyrme field has a polyhedral shape consisting of a ring of eight 
pentagons capped by squares above and below.

In a sense, the $B=7$ case is similar to the case $B=6$,
but the \sk\ has a dodecahedral shape. A dodecahedron is a
ring of ten pentagons capped by pentagons above and below.
Among the degree 7 rational maps with $D_{5d}$ symmetry 
\begin{equation}
R(z)=\frac{z^5-a}{z^2(az^5+1)}, 
\end{equation} 
the one with icosahedral symmetry
has $a=-1/7$ (not $a=-3$ as stated in \cite{HMS}). The Wronskian is then 
proportional to $z(z^{10}+11z^5-1)$, the face polynomial of a dodecahedron. 
In another orientation, tetrahedral symmetry $T$ is manifest. There is a 
one-parameter family of maps with the symmetries (\ref{z2b}) and (\ref{s3}),
\begin{equation}
R(z)=\frac{bz^6-7z^4-bz^2-1}{z(z^6+bz^4+7z^2-b)}
\label{g7}
\end{equation} 
where $b$ is complex. 
For real $b$, the symmetry extends to $T_h$ and for $b$ imaginary it
extends to $T_d.$ When $b=0$ there is cubic symmetry $O_h$, and when $b=\pm
7/\sqrt{5}$ there is icosahedral symmetry $Y_h$. Using (\ref{g7}) in 
our ansatz, one finds the minimal energy at 
$b=\pm 7/\sqrt{5}$, which gives a dodecahedral \Sk\ field, with energy
$E=1.107$. This is particularly close to the energy of the true
solution. There is a saddle point at $b=0$ with a cubic shape.

The $B=8$ \sk\ has symmetry $D_{6d}$, as do the rational maps
\begin{equation}
R(z)=\frac{z^6-a}{z^2(az^6+1)}.
\label{g8}
\end{equation} This time the minimal energy  
is $E=1.118$ when $a=0.14.$ The shape is now a ring
of twelve pentagons capped by hexagons above and below.


\begin{figure}[hp]
\unitlength1cm
\begin{picture}(12,10)
\epsfxsize=12cm
\epsffile{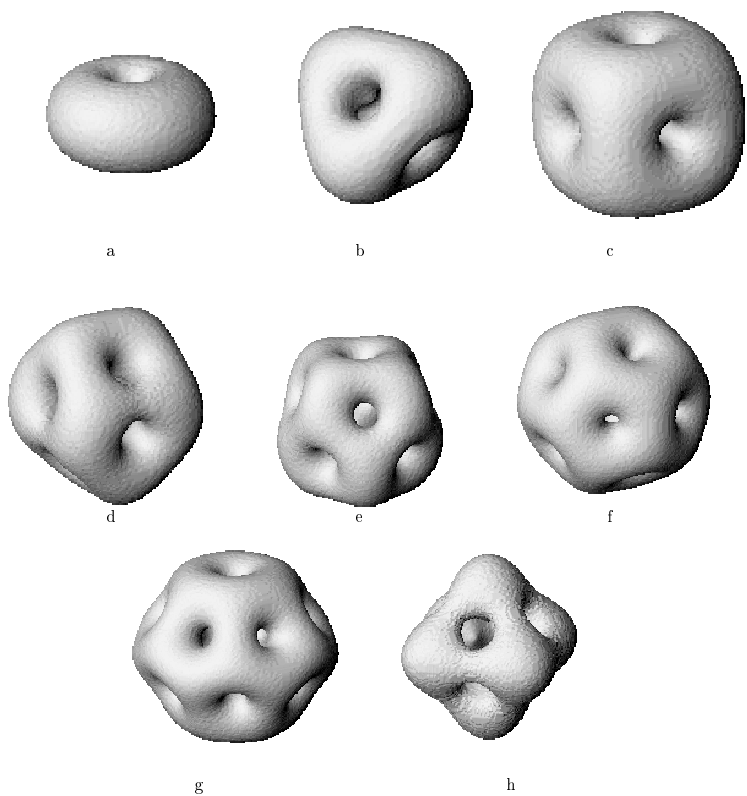}
\end{picture}
\caption{Surfaces of constant baryon density for the following approximate
Skyrmions, constructed using rational maps: a) $B=2$ torus; 
b) $B=3$ tetrahedron; c) $B=4$ cube; d) $B=5$ with $D_{2d}$ symmetry;
e) $B=6$ with $D_{4d}$ symmetry; f) $B=7$ dodecahedron; 
g) $B=8$  with $D_{6d}$ symmetry; h) $B=5$ octahedron (saddle point).}
\end{figure}

\vbox{
\begin{center}
\begin{tabular}{|c|c|c|c|c|} \hline
B& $\I$ & APPROX & TRUE  & SYM\\
\hline
1 &  1.00 & 1.232 & 1.232  & $O(3)$ \\
2 &  5.81 & 1.208 & 1.171  & $O(2)\times\Z_2$ \\
3 &  13.58 & 1.184 & 1.143   & $T_{d}$ \\
4 & 20.65 & 1.137 & 1.116   & $O_h$ \\
5 & 35.75 & 1.147 & 1.116   & $D_{2d}$ \\
6 & 50.76 & 1.137& 1.109   & $D_{4d}$ \\
7 & 60.87 & 1.107 & 1.099   &  $Y_{h}$ \\
8 & 85.63 & 1.118 & 1.100   &  $D_{6d}$ \\
\hline
5* & 52.05 & 1.232 & 1.138 &  $O_{h}$ \\
\hline
\end{tabular}
\end{center}
{\bf Table 1} : The energies of approximate
Skyrmions generated from rational maps, and of true
Skyrmions. The table gives the value of the angular integral $\I$, and
the associated Skyrme field energy (APPROX), together with the energy
of the true solution (TRUE), as determined in refs. \cite{HMS,BS2},
and the symmetry (SYM) of the solution.  A $*$
denotes a saddle point configuration.}

\section{Multi-shell Rational Maps}
\news

The minimal energy solution of the Skyrme equation (\ref{sky}) with
infinite baryon number is a three-dimensional cubic
crystal. It is obtained by relaxing a face-centred cubic array of
Skyrmions \cite{CJJ}. For finite, increasing $B$,
the single-shell polyhedral structures we have discussed so far are
therefore unlikely to remain
the minimal energy solutions. Skyrmions will probably look
more like part of the crystal. An approximate construction of
Skyrmions as part of the crystal was carried out by Baskerville, for
some special values of $B$, but
the resulting energies were rather high \cite{Ba}. Here we try
a rational map ansatz with a two-shell structure. This is easily
generalized to a multi-shell structure. The connection with the
crystal will emerge below.

The simplest version of this generalized ansatz is (\ref{ansatz})
itself, with the profile function $f(r)$ having boundary values $f(0)=
2\pi \ , \ f(\infty)= 0$. However, this does not give a low energy. More
promising is to use two different rational maps, $R_1(z)$ of degree
$N_1$ for the inner shell, and $R_2(z)$ of degree $N_2$ for the outer shell. 
Let $r_0 > 0$ denote the radius where the inner and outer shells
join. The ansatz is now

\begin{equation}
U(r,z)= \left\{ \begin{array}{ll}
\exp(if_1(r)\ {\widehat{\bf n}_{R_1(z)}} \cdot\mbox{\bm $\sigma$}) & 
\mbox{$0 \le r \le r_0$}, \\
{\exp(if_2(r)\ {\widehat{\bf n}_{R_2(z)}} \cdot\mbox{\bm $\sigma$})} &
\mbox{$r_0 \le r$},
\end{array}\right. 
\label{twoshell}
\end{equation}
where the profiles $f_1(r)$ and $f_2(r)$ have boundary values
$f_1(0)=2\pi$, $f_1(r_0)=f_2(r_0)=\pi$, $f_2(\infty)=0$. The field is
continuous at $r=r_0$, but derivatives jump there. Note that $U=1$ at
the centre.

The baryon number of the Skyrme field (\ref{twoshell}) is
easily seen to be $B=N_1 + N_2$. Its energy is the obvious 
generalization of (\ref{energy4}),

\begin{eqnarray}
E&=&4\pi\int_0^{r_0} \bigg(
r^2 {f_1'}^2 +2N_1({f_1'}^2+1)\sin^2 {f_1}+{\I}_1\frac{\sin^4
{f_1}}{r^2}\bigg) \ dr \nonumber \\
&& +4\pi\int_{r_0}^{\infty} \bigg(
r^2 {f_2'}^2+2N_2({f_2'}^2+1)\sin^2 {f_2}+{\I}_2\frac{\sin^4
{f_2}}{r^2}\bigg) \ dr.
\label{energy5}
\end{eqnarray}
There is considerably more choice than before in how to minimize
this. One should consider all pairs $N_1,N_2$ whose sum is $B$, then
find rational maps that minimize 
${\I}_1$ and ${\I}_2$, and finally find the profiles 
$f_1(r)$ and $f_2(r)$, allowing $r_0$ to be variable. We have not
carried out such a systematic analysis. Instead, we have considered
those pairs of maps $R_1(z)$ and $R_2(z)$ which have a high degree of
symmetry, and which appear to fit well together. Our aim is to obtain
a field with a low, but possibly not optimal, value of the energy
(\ref{energy5}). We have then relaxed this field numerically to find a
true solution of the Skyrme equation, usually with the same
symmetry. We have used this approach for
baryon numbers $B=12,13 \ {\rm and} \ 14$. In each of these cases
there is the possibility of a cubically symmetric solution, rather
similar to part of the Skyrme crystal. We describe these in turn.\\

\subsection*{$B=12$}

There are various attractive choices for $N_1$ and $N_2$.
Choosing $N_1 = N_2 =6$, with the rational map (\ref{g6}), gives a
rather low symmetry. More interesting is $N_1 = 3$ and $N_2 =9$, where
there are tetrahedrally symmetric maps. However, the most successful
choice is $N_1 = 5$ and $N_2 =7$. One could use the optimal
single-shell maps given earlier (for $B = 5,7$), but they have low
combined symmetry. Better is to combine the maps with cubic
symmetry mentioned earlier
\begin{equation}
R_5(z)=\frac{z(z^4-5)}{-5z^4+1} \hskip 1cm \mbox{,} \hskip 1cm
R_7(z)=\frac{-7z^4-1}{z^3(z^4+7)}.
\label{g12}
\end{equation}
These maps both have the tetrahedral symmetries (\ref{z2b}) and
(\ref{s3}), as well as the $90^{\circ}$ rotation symmetry $R(iz) =
iR(z)$.

We have calculated the optimal profile functions and optimal $r_0$
for this pair of maps, obtaining an energy $E=1.30$. Then, with this as
a starting point, we have numerically relaxed the field to obtain a
cubically symmetric, smooth solution of the Skyrme equation with
energy $E=1.15$. 
Its shape is shown in Figure 2a). Note that the Figure
does not exhibit a two-shell structure. After relaxation, the
inner and outer shells coalesce. We shall return to this below.

Battye and Sutcliffe have also studied the $B=12$ Skyrmion
\cite{BS4}. They have found the optimal single-shell rational map
to use in (\ref{ansatz}). This map has only tetrahedral symmetry $T_d$, and
gives an energy $E=1.102$. They also relax their field to seek a true
solution. This has the same tetrahedral symmetry, and energy $E=1.086$.
The cubically symmetric solution is not the true Skyrmion, but probably a
saddle point.

\subsection*{$B=13$}

To construct a cubically symmetric $B=13$ Skyrme field one might try a
three-shell structure, with baryon numbers 1+7+5 from the centre
outward, combining the map $R_1(z) = z$ with the maps in 
(\ref{g12}). However, this cannot easily be implemented, because of
the large size and energy of the initial configuration. Instead, we have
constructed the solution, which resembles part
of the Skyrme crystal, by relaxing a configuration made from
a single Skyrmion and twelve nearest neighbours. It is shown in
Figure 2b), and has energy $E=1.09$.

Battye and Sutcliffe have also investigated the $B=13$ case, using the
single-shell ansatz. The rational map minimizing $\I$ has 
$O$ symmetry. However, there is an $O_h$ symmetric map with a slightly
larger value of $\I$, and using this gives a field which looks
almost identical to Figure 2b). We conclude that there is an $O_h$
symmetric $B=13$ solution, with $U=-1$ at the centre, 
which can be found starting in several ways. However, it
appears that the solution with just $O$ symmetry is the true Skyrmion.

\subsection*{$B=14$}

Again we seek a cubically symmetric solution. We do this by taking a
two-shell ansatz, using the dodecahedral rational maps of degree
7. The inner shell map is (\ref{g7}) with $b= 7/\sqrt{5}$, the
outer shell uses the same map with $b= -7/\sqrt{5}$. Together, these
maps (at different radii) possess only $T_h$ symmetry, but if they can
be made to coalesce, then there is cubic symmetry, because a 
$90^{\circ}$ rotation transforms one into the other. 
Optimizing the profile functions gives an
energy $E=1.39$. Further relaxation produces a solution with $T_h$ 
symmetry, and nearly cubically symmetric, with $E=1.14$. Its form
is shown in Figure 2c). This again looks like part of the Skyrme
crystal; this time what one would obtain by taking the six nearest
neighbours and eight next-nearest neighbours surrounding a hole in 
the ``face centered'' cubic array of Skyrmions, and relaxing the field.

The optimal single-shell structure with $B=14$ is quite different 
\cite{BS4}. The
rational map has only $D_2$ symmetry, and gives an energy $E=1.103$.
The field also differs because $U=-1$ at the centre.
Relaxation of the solution will give a lower energy, 
but this has not yet been done.

\section{Interpretation of the Two-shell Ansatz}
\news

For certain profile functions $f_1$ and $f_2$, 
the two-shell rational map ansatz describes a Skyrmion of baryon
number $N_1$ inside a Skyrmion of baryon number $N_2$, approximately. 
However, as the
field relaxes, it changes its character. Consider a
radial line ($z$ fixed), and the field values $U$ at the points along
it where $f_1 = 3\pi/2$ and where $f_2 = \pi/2$. If these values are
close, then the field between can be relaxed to be approximately
constant, which makes the energy low in this direction. Conversely, if
the field values are antipodal (on $SU(2)$), then the field gradient
between them, and hence the energy, is large in this direction. In
fact the winding of the field along this radial line indicates that
there is a $B=1$ Skyrmion in this direction.

Now, antipodal field values occur on this line if $R_1(z)=R_2(z)$. (The
rational map values are the same, but $\sin f_1 = -1 \ , \ \sin f_2 =
1$.) Thus the two-shell rational map ansatz produces a configuration
which can be interpreted as a superposition of $B=1$ Skyrmions
located at $r=r_0$ and in those directions $z$ which solve the
equation
\begin{equation}
R_1(z)=R_2(z).
\label{equalmap}
\end{equation}
Writing $R_1(z)= \frac{p_1(z)}{q_1(z)}$ and $R_2(z)=
\frac{p_2(z)}{q_2(z)}$, this becomes
\begin{equation}
p_1(z)q_2(z)-p_2(z)q_1(z)=0
\label{polyeq}
\end{equation}
which is a polynomial equation of degree $N_1 + N_2$, with $N_1 + N_2$
solutions. So the number of Skyrmions one finds by solving
(\ref{polyeq}) is precisely the total baryon number. 
The relative orientation of these Skyrmions has not yet been determined.

Eq. (\ref{polyeq}) has a particularly symmetric form for
maps we have been considering. For $B=12$ it reduces to 
\begin{equation}
z^{12} - 33z^8 - 33z^4 + 1=0,
\label{edge}
\end{equation}
the Klein polynomial for the edges of a cube. For $B=14$ it reduces to
\begin{equation}
z(z^4 -1)(z^8 + 14z^4 + 1)=0,
\label{vertface}
\end{equation}
the product of the Klein polynomials for the faces and vertices of a
cube (one root is $z=\infty$). 
This is what one anticipates based on the analogy with
the Skyrme crystal.

\begin{figure}[hp]
\unitlength1cm
\begin{picture}(9,3)
\epsfxsize=12cm
\epsffile{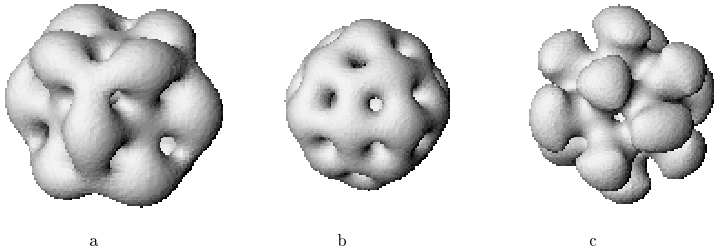}
\end{picture}
\caption{Surfaces of constant baryon density for the following 
solutions: a) $B=12$ with cubic symmetry; 
b) $B=13$ with cubic symmetry; c) $B=14$ two-shell with near cubic symmetry.} 
\end{figure}

\section{Conclusions}
We have discussed an ansatz for Skyrme fields, based on rational maps, which 
allows the construction of good approximations to several Skyrmions.
We have also discussed a two-shell rational map ansatz as an approach to 
construct multi-shell Skyrmion solutions. 
These ans\"atze are good starting points to construct solutions with
certain symmetries. We have studied other examples of the two-shell
rational map ansatz than the ones described here and in most cases the 
configuration relaxes to a single shell solution. Two-shell solutions were
found only for $B=14$, as we have shown, and also when using a rational map 
of degree $17$, together with another of lower degree. To construct two-shell 
solutions with a relatively low energy, 
the outside shell must be large enough to contain a smaller shell inside it.
Although the solutions we have found using the two-shell ansatz are not
minimal energy Skyrmions, their energies are not much greater, and we
believe that for higher baryon numbers the
minimal energy solutions will exhibit a multi-shell structure.

 

\section*{Acknowledgements}

We thank Richard Battye, Paul Sutcliffe and Wojtek Zakrzewski
for useful discussions.

\end{document}